\documentclass[12pt]{elsart}
\usepackage{feynmp,epsfig,graphics}
\def\be{\begin{eqnarray}}
\def\ee{\end{eqnarray}}
\def\bc{\begin{center}}
\def\ec{\end{center}}

\newcommand{\tr}{\rm tr \,}

\begin{document}
GSI-Preprint-2003-20; hep-ph/0307133
\begin{frontmatter}
\title{On heavy-light meson resonances and chiral symmetry }

\author[NBI]{E.E. Kolomeitsev}
\author[GSI,TU]{and M.F.M. Lutz}
\address[NBI]{The Niels Bohr Institute\\ Blegdamsvej 17, DK-2100 Copenhagen, Denmark}
\address[GSI]{Gesellschaft f\"ur Schwerionenforschung (GSI)\\
Planck Str. 1, 64291 Darmstadt, Germany}
\address[TU]{Institut f\"ur Kernphysik, TU Darmstadt\\
D-64289 Darmstadt, Germany}
\begin{abstract}
We study heavy-light meson resonances with quantum numbers $J^P\!=\!0^+$ and $J^P=1^+$ in terms
of the non-linear chiral SU(3) Lagrangian. At leading order a parameter-free
prediction is obtained for the scattering of Goldstone bosons off
heavy-light pseudo-scalar and vector mesons once we insist on approximate crossing symmetry
of the unitarized scattering amplitude. The recently announced narrow open charm states
observed by the BABAR and CLEO collaborations are reproduced. We suggest the existence of
states that form an anti-triplet and a sextet representation of the SU(3) group. In particular,
so far unobserved narrow isospin-singlet states with negative strangeness are
predicted. The open bottom states with $(I,S)=(0,-1)$ are anticipated
at 5761 MeV ($J^P\!=\!0^+$) and 5807 MeV ($J^P\!=\!1^+$). For the anti-triplet states
our results differ most significantly from predictions that are based on a linear realization
of the chiral SU(3) symmetry in the open bottom sector. Strongly bound $0^+$-and $1^+$-states
with $(I,S)=(0,1)$ at 5643 MeV and 5690 MeV are predicted.
\end{abstract}
\end{frontmatter}

\section{Introduction}

In  a recent work \cite{LK03} it was demonstrated that
chiral SU(3) symmetry predicts parameter-free $J^P\!=\!1^+$ light meson resonances.
It was observed that the resonance states turn into
bound states in the heavy SU(3) limit with $m_{\pi,K,\eta}\simeq 500$ MeV
but disappear altogether in the light SU(3) limit with
$m_{\pi,K,\eta}\simeq 140$ MeV. In earlier works \cite{Rupp86,WI90,JPHS95,OOP99,NVA02,NP02}
similar results were obtained for light meson resonances with $J^P\!=\!0^+$ quantum numbers.
In view of the apparent success of the chiral coupled-channel dynamics to
predict the existence of a wealth of meson resonances in the ($u,d,s$)-sector of QCD it
is interesting to study whether the same mechanism is able to predict heavy-light meson
resonances, i.e. meson resonances with open charm or bottom.
Recently a new narrow state of mass 2.317 GeV that decays into $D^+_s\,\pi^0$ was announced
\cite{BaBar}. This result was confirmed \cite{CLEO} and a second narrow state of mass
2.46 GeV decaying into $D_s^* \pi^0$ was observed. Such states were first predicted in
\cite{NRZ93,BH94} based on the spontaneous breaking of chiral symmetry. Since one expects
from such studies \cite{NRZ93,BH94,BEH03,NRZ03} the existence of further so far unobserved
states it is important to study the heavy-light resonances in great detail \cite{all-papers}.

The theoretical predictions \cite{NRZ93,BH94,BEH03,NRZ03} rely on the chiral quark
model which implies a linear realization of the chiral SU(3) symmetry of QCD.
The linear realization predicts the $0^+,1^+$ resonance
states to form an anti-triplet representation of the SU(3) group. This is not necessarily so in the
non-linear representation of the SU(3) group.  If one insists on a non-linear
realization of the chiral SU(3) group no a priori prediction can be made for the existence
of chiral partners of any given state. For instance in \cite{LK03} is was shown that
the light $1^+$ spectrum predicted by the non-linear chiral representation forms two
degenerate octets and an additional singlet consistent with the empirical spectrum.
Analogous results were obtained for the light scalar mesons \cite{Rupp86,WI90,JPHS95,OOP99,NVA02,NP02}

In this work we apply the $\chi$-BS(3) approach developed
originally for meson-baryon scattering \cite{LK00,LK01,LH02,LK02,Granada,Copenhagen} but
recently also applied to meson-meson scattering \cite{LK03}. Using the chiral SU(3)
Lagrangian involving light-heavy $J^P\!=\!0^-$ and $J^P\!=\!1^-$ fields that transform non-linear under
the chiral SU(3) group a coupled-channel description of the meson-meson scattering in the open
charm and bottom sector is developed. The possible importance of coupled-channel dynamics
for the heavy-light meson states was emphasized recently \cite{BR03}.
The major result of our work is the prediction that there
exist states with $J^P\!=\!0^+,1^+$ quantum numbers forming anti-triplet and sextet
representations of the SU(3) group. This differs from the results
implied by the linear realization of the chiral SU(3) symmetry leading to
anti-triplet state only. Our result
suggests the existence of $J^P\!=\!0^+, 1^+$ states with unconventional quantum numbers
$(I,S)=(1,1)$ and $(I,S)=(0,-1)$. A particular result concerns
the 'heavy' SU(3) limit with $m_{\pi ,\eta, K} \sim 500$ MeV  and $M_{D}\sim 1800$ MeV in which
the chiral coupled-channel dynamics predicts anti-triplet bound states rather than
resonance states only. In the 'light' SU(3) limit with $m_{\pi ,\eta, K} \sim 140$ MeV  and
$M_{D}\sim 1800$ MeV we do not find anymore resonances or bound states in the $J^P\!=\!0^+, 1^+$
sectors. Using physical mass parameters we predict narrow $J^P\!=\!0^+, 1^+$ states in the
$(I,S)= (0,1),(\frac{1}{2},0),(0,-1)$ channels with open charm and open bottom. The open charm
$(0,1)$ states established by the BABAR \cite{BaBar} and CLEO \cite{CLEO} collaborations
are recovered within 15 MeV accuracy. We identify the observed $D(2420)$ resonance with
$J^P=1^+$ and $(I,S)=(\frac{1}{2},0)$ to be a member of the
sextet. This suggests the existence of isospin zero $\bar K\,D(1867)$-and
$\bar K\,D(2008)$-bound states.

\section{Chiral coupled-channel dynamics: the $\chi$-BS(3) approach}

The starting point to study the scattering of Goldstone bosons off
heavy-light mesons is the chiral SU(3) Lagrangian.
We identify the leading-order Lagrangian
density \cite{Wein-Tomo,Wise92,YCCLLY92,BD92} describing the interaction of
Goldstone bosons with pseudo-scalar and vector mesons,
\begin{eqnarray}
{\mathcal L}(x) &=&
-\frac{1}{8\,f^2}\,\tr  \Big[
H_{\phantom{\mu}}(x)\, (\partial^\nu H^\dagger_{\phantom{\mu}}(x) )
-(\partial^\nu H_{\phantom{\mu}}(x))\,H^\dagger_{\phantom{\mu}}(x)  \Big]
\,[\phi (x) , (\partial_\nu\,\phi(x))]_-
\nonumber\\
&-& \frac{1}{8\,f^2}\,\tr
\Big[
H^\mu(x)\, (\partial^\nu H^\dagger_\mu(x) )-(\partial^\nu H^\mu(x))\,H^\dagger_\mu(x)  \Big]
\,[\phi (x) , (\partial_\nu\,\phi(x))]_- \,,
\label{WT-term}
\end{eqnarray}
where $\phi$ is the Goldstone bosons field  and $H$
and $H_\mu$ are massive pseudo-scalar and vector-meson fields. The parameter
$f$ in (\ref{WT-term}) is known from the weak decay process of the
pions. We use $f= 90$ MeV through out this work.
Since we will assume perfect
isospin symmetry it is convenient to decompose
the fields into their isospin multiplets. The fields can be written in terms of
isospin multiplet fields like $K =(K^{(+)},K^{(0)})^t $ and $D=(D^{(+)},D^{(0)})^t$,
\begin{eqnarray}
&& \phi = \tau \cdot \pi (140)
+ \alpha^\dagger \cdot  K (494) +  K^\dagger(494) \cdot \alpha
+ \eta(547)\,\lambda_8\,,
\nonumber\\
&& H_{\phantom{\mu}}  = {\textstyle{1\over \sqrt{2}}}\,\alpha^\dagger \cdot D_{\phantom{\mu}}(1867)
- {\textstyle{1\over \sqrt{2}}}\,D^t_{\phantom{\mu}}(1867)\cdot \alpha   +  i\,\tau_2\,D_{\phantom{\mu}}^{(s)}(1969) \,,
\nonumber\\
&& H_\mu = {\textstyle{1\over \sqrt{2}}}\,\alpha^\dagger \cdot D_\mu(2008)
- {\textstyle{1\over \sqrt{2}}}\,D^t_\mu(2008)\cdot \alpha   +  i\,\tau_2\,D^{(s)}_{\mu}(2110) \,,
\nonumber\\
&& \alpha^\dagger = {\textstyle{1\over \sqrt{2}}}\,(\lambda_4+i\,\lambda_5 ,\lambda_6+i\,\lambda_7 )\,,\qquad
\tau = (\lambda_1,\lambda_2, \lambda_3)\,,
\label{}
\end{eqnarray}
where the matrices $\lambda_i$ are the standard Gell-Mann generators of the SU(3) algebra.
The numbers in the brackets recall the approximate masses of the fields in units of MeV.

\begin{table}
\tabcolsep=-.1mm
\begin{tabular}{||cccccccccccc||}
\hline\hline
\multicolumn{3}{||c|}{$(\frac{1}{2},+2)$} &
\multicolumn{3}{c|}{$(0,+1)$} &
\multicolumn{3}{c|}{$(1,+1)$} &
\multicolumn{3}{c|}{$(\frac12,0)$}
\\ \hline
\multicolumn{3}{||c|}{$(D_s\,K)$} &
\multicolumn{3}{c|}{$\left(\begin{array}{c} ({\textstyle{1\over
\sqrt{2}}}\,D^t \, i\,\sigma_2\, K)\\ (D_s\,\eta )
\end{array}\right)$}&
\multicolumn{3}{c|}{$\left(\begin{array}{c} (D_s \, \pi)\\ (
{\textstyle{1\over
\sqrt{2}}}\,D^t \, i\,\sigma_2\,\sigma\, K)
\end{array}\right)$} &
\multicolumn{3}{c|}{$\left(\begin{array}{c} ({\textstyle{1\over
\sqrt{3}}}\,\pi \cdot
\sigma\, D)\\ ( \eta \,D) \\ (D_s\,i\,\sigma_2\,\bar K^t)
\end{array}\right)$}
\\\hline\hline
\multicolumn{4}{||c|}{\phantom{xxxxxx}$(\frac32,0)$\phantom{xxxxxx}} &
\multicolumn{4}{c|}{\phantom{xxxxxx}$(0,-1)$\phantom{xxxxxx}} &
\multicolumn{4}{c|}{$(1,-1)$}
\\\hline
\multicolumn{4}{||c|}{$( \pi \cdot T\,D)$} &
\multicolumn{4}{c|}{$( {\textstyle{1\over \sqrt{2}}}\,\bar K
\,D)$} &
\multicolumn{4}{c|}{$({\textstyle{1\over \sqrt{2}}}\,\bar K
\,\sigma\,D)$}
\\\hline\hline
\end{tabular}
\caption{The definition of coupled-channel states with $(I,S)$.}
\label{tab:states}
\end{table}

The scattering problem decouples into seven orthogonal
channels specified by isospin ($I$) and strangeness ($S$) quantum numbers,
\begin{eqnarray}
(I,S)= (({\textstyle{1\over{2}}},+2), (0,+1), (1, +1), ({\textstyle{1\over{2}}}, 0),
({\textstyle{3\over{2}}}, 0), (0,-1), (1,-1))\,.
\label{lwt}
\end{eqnarray}
In Tab. \ref{tab:states} the channels that contribute in a given sector $(I,S)$ are listed.
Heavy-light meson resonances with quantum numbers $J^P\!=\!0^+$ and $J^P\!=\!1^+$ manifest
themselves as poles in the s-wave scattering amplitudes, $M^{(I,S)}_{J^P}(\sqrt{s}\,)$, which
in the $\chi-$BS(3) approach \cite{LK02,LK03} take the simple form
\begin{eqnarray}
&&  M^{(I,S)}_{J^P}(\sqrt{s}\,) = \Big[ 1- V^{(I,S)}(\sqrt{s}\,)\,J^{(I,S)}_{J^P}(\sqrt{s}\,)\Big]^{-1}\,
V^{(I,S)}(\sqrt{s}\,)\,.
\label{final-t}
\end{eqnarray}
The effective interaction kernel $V^{(I,S)}(\sqrt{s}\,)$ in (\ref{final-t}) is determined by
the leading order chiral SU(3) Lagrangian (\ref{WT-term}),
\begin{eqnarray}
&& V^{(I,S)}(\sqrt{s}\,) = \frac{C^{(I,S)}}{8\,f^2}\, \Big(
3\,s-M^2-\bar M^2-m^2-\bar m^2
\nonumber\\
&& \qquad \qquad \qquad \qquad \qquad -\frac{M^2-m^2}{s}\,(\bar M^2-\bar m^2)\Big) \,,
\label{VWT}
\end{eqnarray}
where $(m,M)$ and $(\bar m, \bar M)$ are the masses of initial and final mesons. We use
capital $M$ for the masses of heavy-light mesons and small $m$ for the masses of the Goldstone
bosons. The matrix of coefficients $C^{(I,S)}$ that characterize the interaction strength in
a given channel is given in Tab. \ref{tab:coeff}.
The s-wave interaction kernels are identical for the two scattering problems considered here.

In contrast the loop functions, diagonal in the coupled-channel space, depend on
whether to scatter Goldstone bosons off pseudo-scalar or vector heavy-light mesons,
\begin{eqnarray}
&& J_{0^+}(\sqrt{s}\,) = I(\sqrt{s}\,)-I(\mu_{0^+}^{(I,S)})\,,
\nonumber\\
&& J_{1^+}(\sqrt{s}\,) =
\Big(1 + \frac{p_{\rm cm}^2}{3\,M^2} \Big)\,
\Big(I(\sqrt{s}\,)-I(\mu_{1^+}^{(I,S)}) \Big)\,,
\nonumber\\
\nonumber\\
&& I(\sqrt{s}\,)=\frac{1}{16\,\pi^2}
\left( \frac{p_{cm}}{\sqrt{s}}\,
\left( \ln \left(1-\frac{s-2\,p_{cm}\,\sqrt{s}}{m^2+M^2} \right)
-\ln \left(1-\frac{s+2\,p_{cm}\sqrt{s}}{m^2+M^2} \right)\right)
\right.
\nonumber\\
&&\qquad \qquad + \left.
\left(\frac{1}{2}\,\frac{m^2+M^2}{m^2-M^2}
-\frac{m^2-M^2}{2\,s}
\right)
\,\ln \left( \frac{m^2}{M^2}\right) +1 \right)+I(0)\;,
\label{i-def}
\end{eqnarray}
where $\sqrt{s}= \sqrt{M^2+p_{cm}^2}+ \sqrt{m^2+p_{cm}^2}$. Note however that
the two loop functions in (\ref{i-def}) differ by a term suppressed with $1/M^2$ only.
A crucial ingredient of the $\chi-$BS(3) scheme is its approximate crossing symmetry guaranteed
by a proper choice of the subtraction scale $\mu_{J^P}^{(I,S)}$. Referring to the detailed discussions
in \cite{LK02,Granada,Copenhagen,LK03} we obtain
\begin{eqnarray}
&& \mu_{0^+}^{(I,0)} =  M_{D(1867)}\,,\quad
\mu_{0^+}^{(I, \pm 1)} = M_{D_s(1969)}\,,\quad
\mu_{0^+}^{(I, 2)} = M_{D(1867)} \,,
\nonumber\\
&& \mu_{1^+}^{(I,0)} =  M_{D(2008)}\,,\quad
\mu_{1^+}^{(I, \pm 1)} = M_{D_s(2110)}\,,\quad
\mu_{1^+}^{(I, 2)} = M_{D(2008)} \,.
\label{mu-def}
\end{eqnarray}
With (\ref{final-t},\ref{VWT},\ref{i-def},\ref{mu-def}) the brief exposition of the $\chi-$BS(3)
approach as  applied to heavy-light meson resonances is completed.

\begin{table}
\tabcolsep=4.4mm
\begin{center}
\begin{tabular}{||c||p{9.0mm}|p{9.0mm}|p{9.0mm}|p{9.0mm}|p{9.0mm}|p{9.0mm}|p{9.0mm}|p{9.0mm}|p{9.0mm}||}
\hline
($I, S$)    & ($\frac{1}{2}, +2$)  & ($0, +1$) & ($1, +1$) &
($\frac{1}{2},0$) & ($\frac{3}{2},0$) & ($0,-1$) &($1,-1$)  \\
 \hline\hline
11 & $-1$ & $\phantom{-}2$ & $\phantom{-}0$& $\phantom{-}2$ & $-1$ & $\phantom{-}1$ & $-1$  \\ 
  \hline
12 &-- & $\phantom{-}\sqrt{3}$ &$\phantom{-}1$ & $\phantom{-}0$ &-- & -- & --  \\ 
  \hline
22 &-- & $\phantom{-}0$& $\phantom{-}0$  & $\phantom{-}0$& -- & -- & --  \\ 
  \hline
13 &-- & -- &-- & $\phantom{-}\sqrt{\frac{3}{2}}$ &-- & -- & --  \\ 
  \hline
23 &-- & -- &-- & $-\sqrt{\frac{3}{2}}$ &-- & -- &-- \\ 
  \hline
33 &-- &-- &-- &$\phantom{-}1$ &-- &--  &--  \\ 
\hline
\end{tabular}
\caption{The coefficients $C^{(I,S)}$ that characterize the  interaction of Goldstone bosons with heavy meson fields
$H$ and $H_\mu$ as introduced in (\ref{lwt}).} \label{tab:coeff}
\end{center}
\end{table}

\section{Results}

To study the formation of meson resonances we generate speed plots as suggested
by H\"ohler \cite{Hoehler:speed}. The speed ${\rm Speed}_{ab}^{(I,S)}(\sqrt{s})$ of a given
channel $a\,b$ is introduced by \cite{Hoehler:speed,speed},
\begin{eqnarray}
&& t^{(I,S)}_{ab}(\sqrt{s}\,)=\frac{1}{8\,\pi \,\sqrt{s}}\,
\Big( p^{(a)}_{cm}\,p_{cm}^{(b)}\Big)^{1/2}\,M_{ab}^{(I,S)}(\sqrt{s}\,)\,,
\nonumber\\
&& {\rm Speed}_{ab}^{(I,S)}(\sqrt{s}\,) = \Big|\sum_{c}\,
 \Big[\frac{d}{d \,\sqrt{s}}\, t_{ac}^{(I,S)}(\sqrt{s}\,)\Big]\,
 \Big(\delta_{cb}+2\,i\,t_{cb}^{(I,S)}(\sqrt{s}\,) \Big)^\dagger
\Big| \,,
\label{def-speed}
\end{eqnarray}
where we give expressions valid in the $0^+$ sector. A corresponding result for the
$1^+$ sector agrees with (\ref{def-speed}) up to a small correction
term vanishing in the heavy-mass limit $M \to \infty $ (see (\ref{i-def})).

\begin{figure}[t]
\begin{center}
\includegraphics[clip=true,width=13cm]{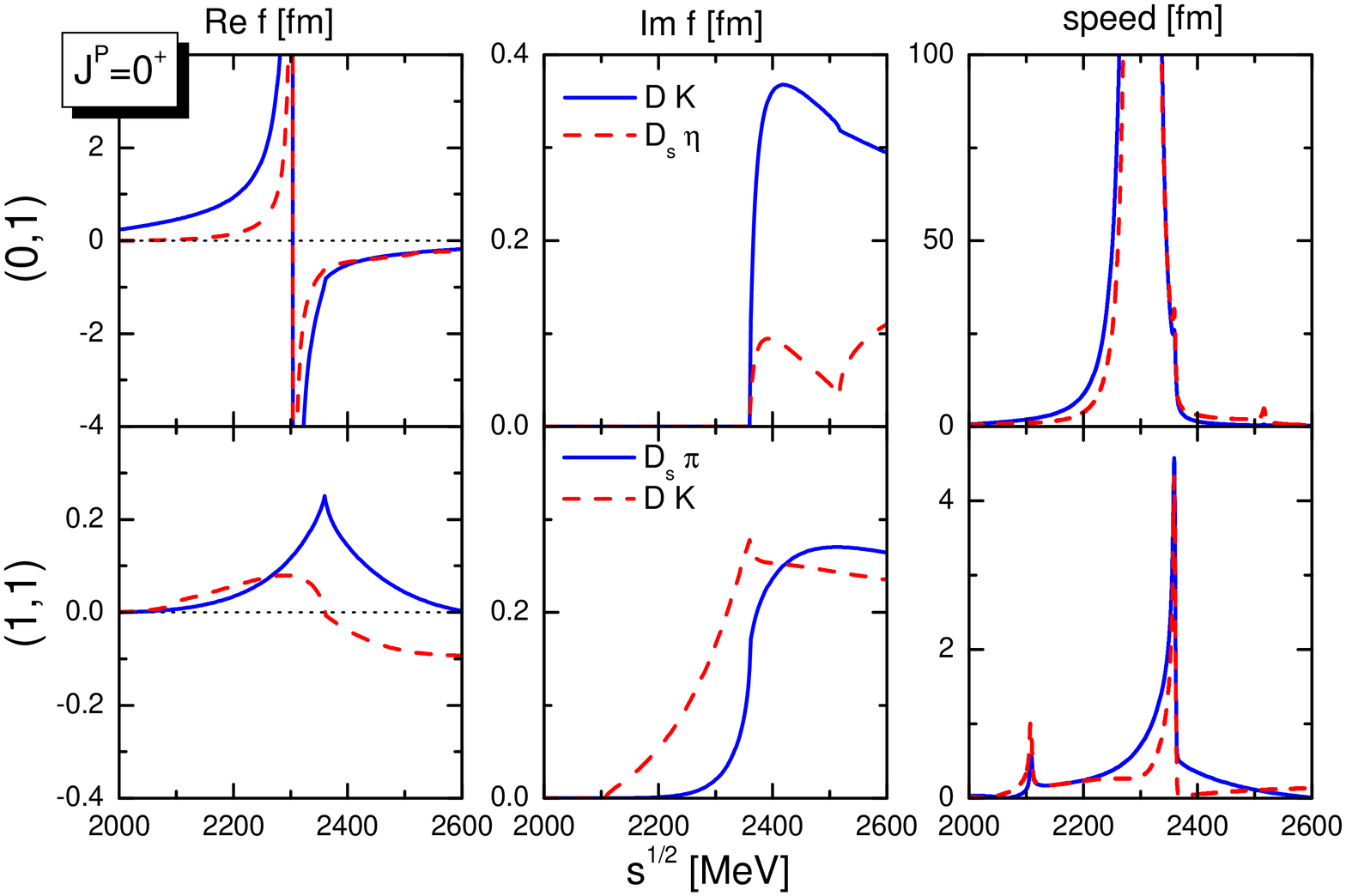}\\
\includegraphics[clip=true,width=13cm]{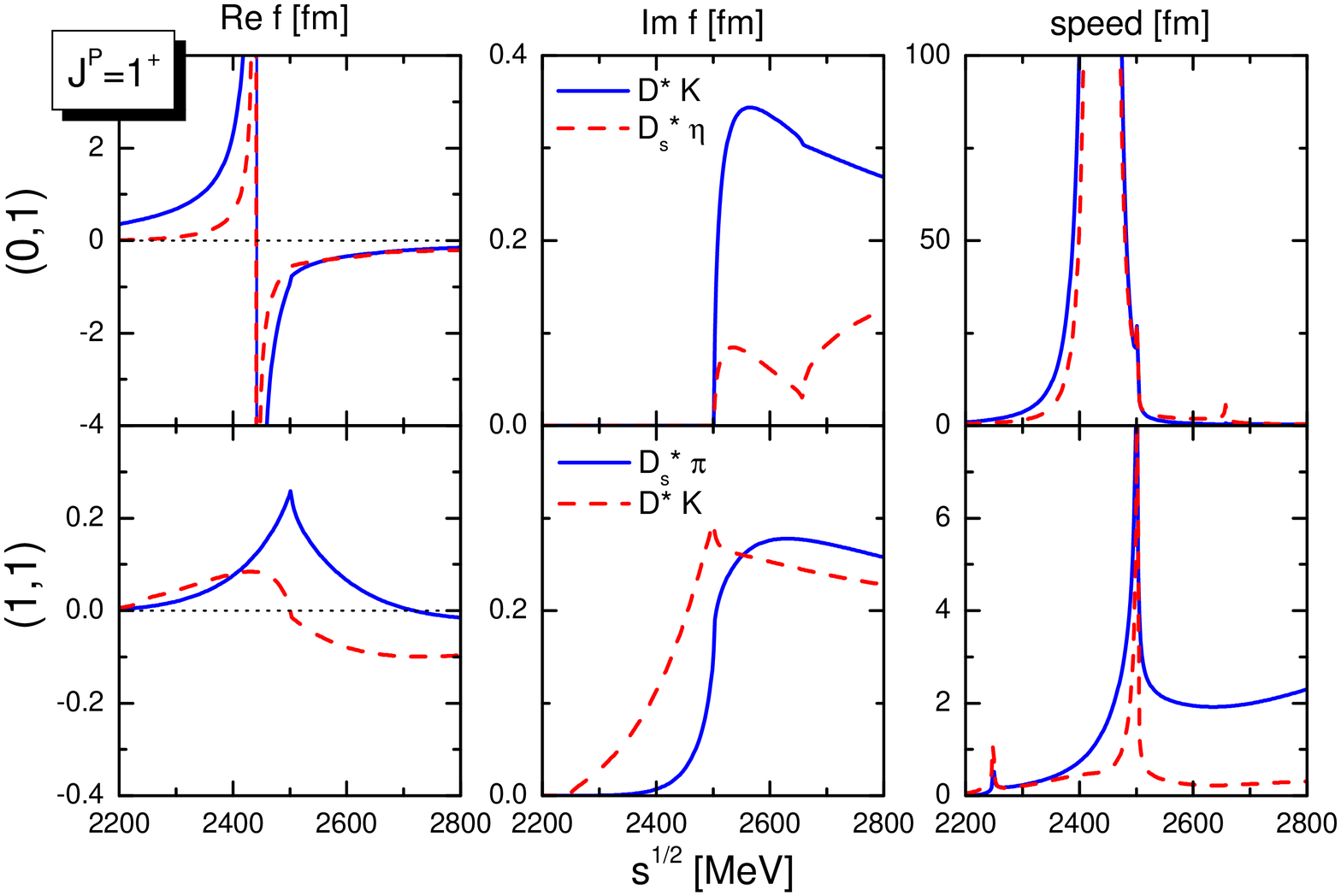}
\end{center}
\caption{Open charm resonances with $J^P\!=\!0^+,1^+$  and $(I,S)=(0,1),(1,1)$ as seen in
the scattering of Goldstone bosons of $D(1867), D_s(1969)$ and
$D(2008), D_s(2110)$ mesons. Shown speed plots together with real and imaginary parts of reduced scattering amplitude,
$f_{ab}$, with $t_{ab}= f_{ab}\,(p^{(a)}_{\rm cm}\,p^{(b)}_{\rm cm})^{1/2}$ (see (\ref{def-speed})).}\label{fig:1}
\end{figure}

\begin{figure}[t]
\begin{center}
\includegraphics[clip=true,width=13cm]{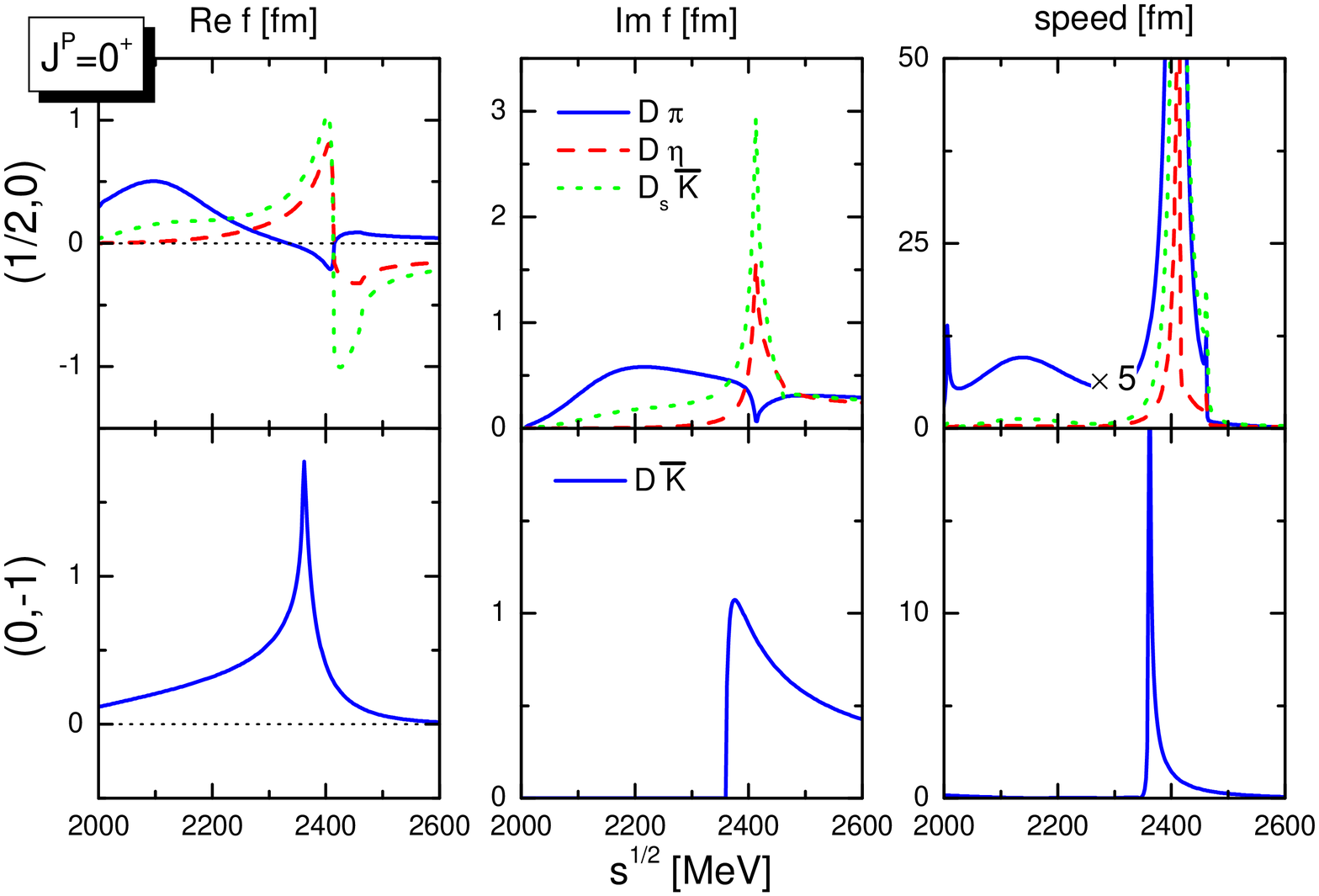}\\
\includegraphics[clip=true,width=13cm]{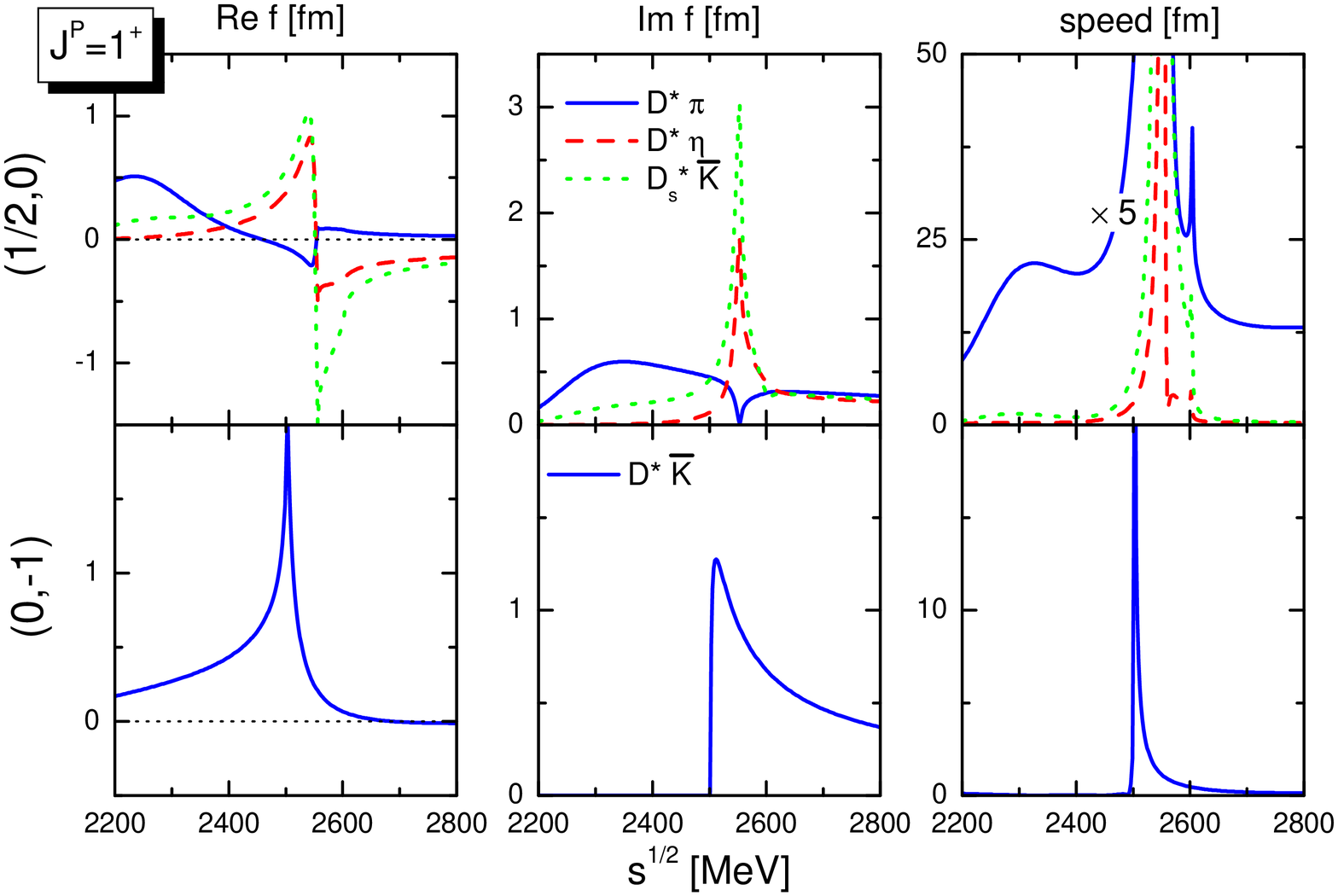}
\end{center}
\caption{Open charm resonances with $J^P\!=\!0^+,1^+$  and $(I,S)=(\frac{1}{2},0),(0,-1)$ as seen in
the scattering of Goldstone bosons of $D(1867), D_s(1969)$ and
$D(2008), D_s(2110)$ mesons. Shown speed plots together with real and imaginary parts of reduced scattering amplitude,
$f_{ab}$, with $t_{ab}= f_{ab}\,(p^{(a)}_{\rm cm}\,p^{(b)}_{\rm cm})^{1/2}$ (see (\ref{def-speed})).}\label{fig:2}
\end{figure}

In order to explore the SU(3) multiplet structure of  the resonance states
we first study the $0^+$ sector in the 'heavy' SU(3) limit \cite{Copenhagen,LK03}
with $m_{\pi ,K,\eta} = 500$ MeV and
$M_{D}=1800$ MeV. In this case we obtain an anti-triplet of mass 2204 MeV with
poles in the $(0,+1),(1/2,0)$ amplitudes. The sextet channel does not show a bound state
signal in this case. However if the attraction is increased slightly by using $f= 80$ MeV
rather than the canonical value $90$ MeV poles at mass 2298 MeV arise
in the $(1,+1),(1/2,0),(0,-1)$ amplitudes.
This finding reflects that the Weinberg-Tomozawa interaction,
\begin{eqnarray}
\bar 3\otimes 8= \bar 3\oplus 6 \oplus \overline{15}
\label{}
\end{eqnarray}
predicts attraction in the anti-triplet and sextet channel but repulsion for the anti-15-plet.
In contrast performing the 'light' SU(3) limit \cite{Copenhagen,LK03} with
$m_{\pi,K,\eta} \sim 140$ MeV together with $M_{D}=1800$ MeV we do not find any signal of a
resonance in any of the channels. Analogous results are found in the $1^+$ sector. If we used
identical masses for the $1^-$ and $0^-$ mesons the differences in the
generated spectra are below 1 MeV.

In Figs. \ref{fig:1},\ref{fig:2} the spectrum as it arises with physical masses is shown.
We predict a bound state of mass 2303 MeV in the $(0,1)$-sector (see Fig. \ref{fig:1}).
According to \cite{BEH03,NRZ03} this state should be identified with a narrow resonance of
mass 2317 MeV
recently observed by the BABAR collaboration \cite{BaBar}. Since we do not consider isospin
violating processes like $\eta \to \pi_0$ the latter state is a true bound state in our
present scheme. Given the fact that our computation is parameter-free this is a remarkable
result. As demonstrated by the real part of the corresponding scattering amplitude
of Fig. \ref{fig:1} the state couples dominantly to the $D\,K$ channel.
In the $(1,+1)$-speeds where we expect a signal from  the sextet
a strong cusp effect at the $K\,D(1867)$-threshold is seen. The large coupling constant to the
$\pi\,D_s(1969)$ channel leads to the broad structure seen in the figure.
Fig. \ref{fig:2} illustrates that in the $(\frac{1}{2},0)$-sector we
predict a narrow state of mass 2413 MeV just below the $\eta\,D(1867)$-threshold and a
broad state of mass 2138 MeV. Modulo some mixing
effects the heavier of the two is part of the sextet the lighter a member of the anti-triplet.
The latter $(\frac{1}{2},0)$-state was expected to have a large branching ratio into the
$\pi \,D(1867)$-channel \cite{BR03,NRZ03}. This is confirmed by our analysis.
Finally in the $(0,-1)$-speed a pronounced cusp effect at the $\bar K\,D(1867)$-threshold
is seen.

\begin{figure}[t]
\begin{center}
\includegraphics[clip=true,width=13cm]{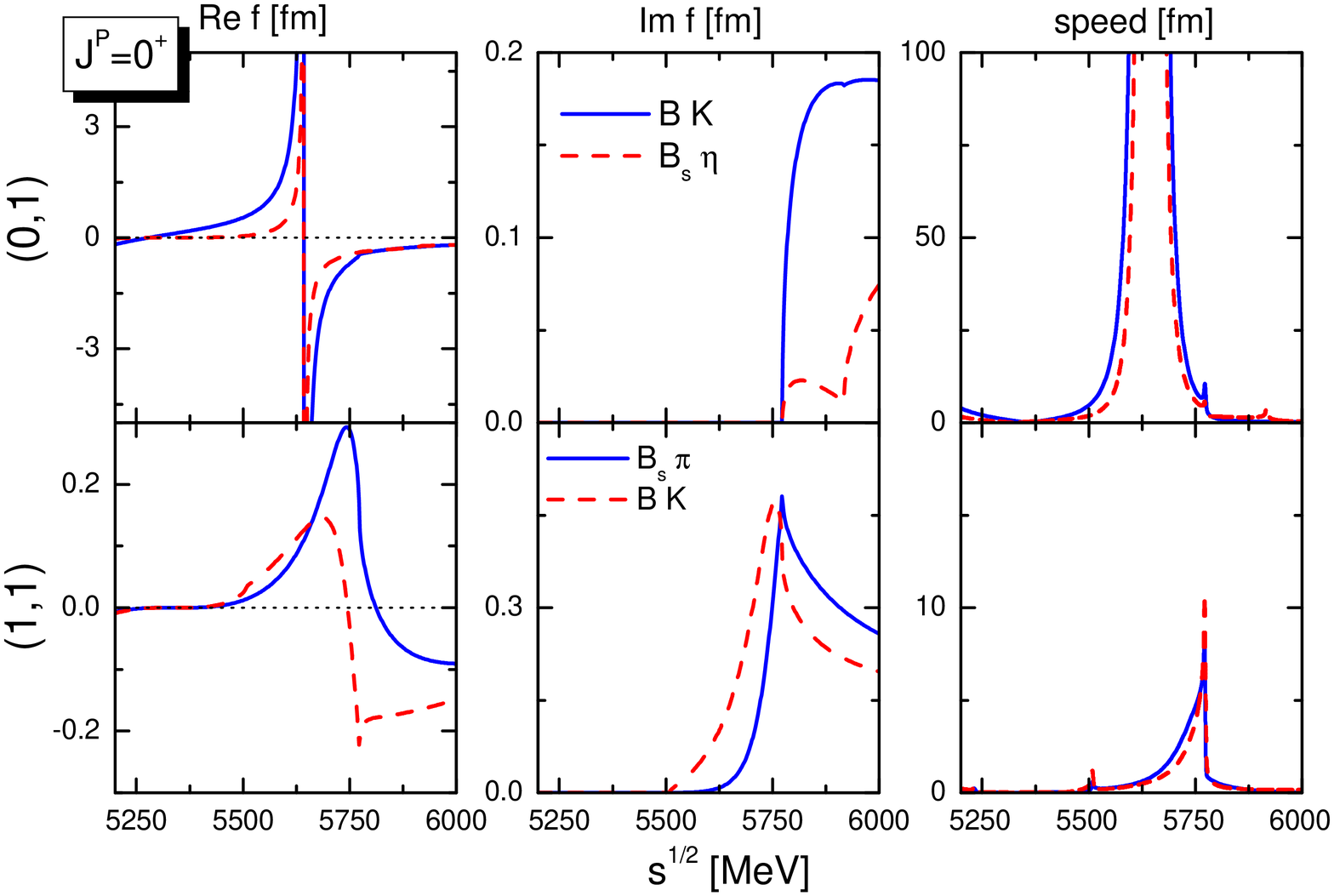}\\
\includegraphics[clip=true,width=13cm]{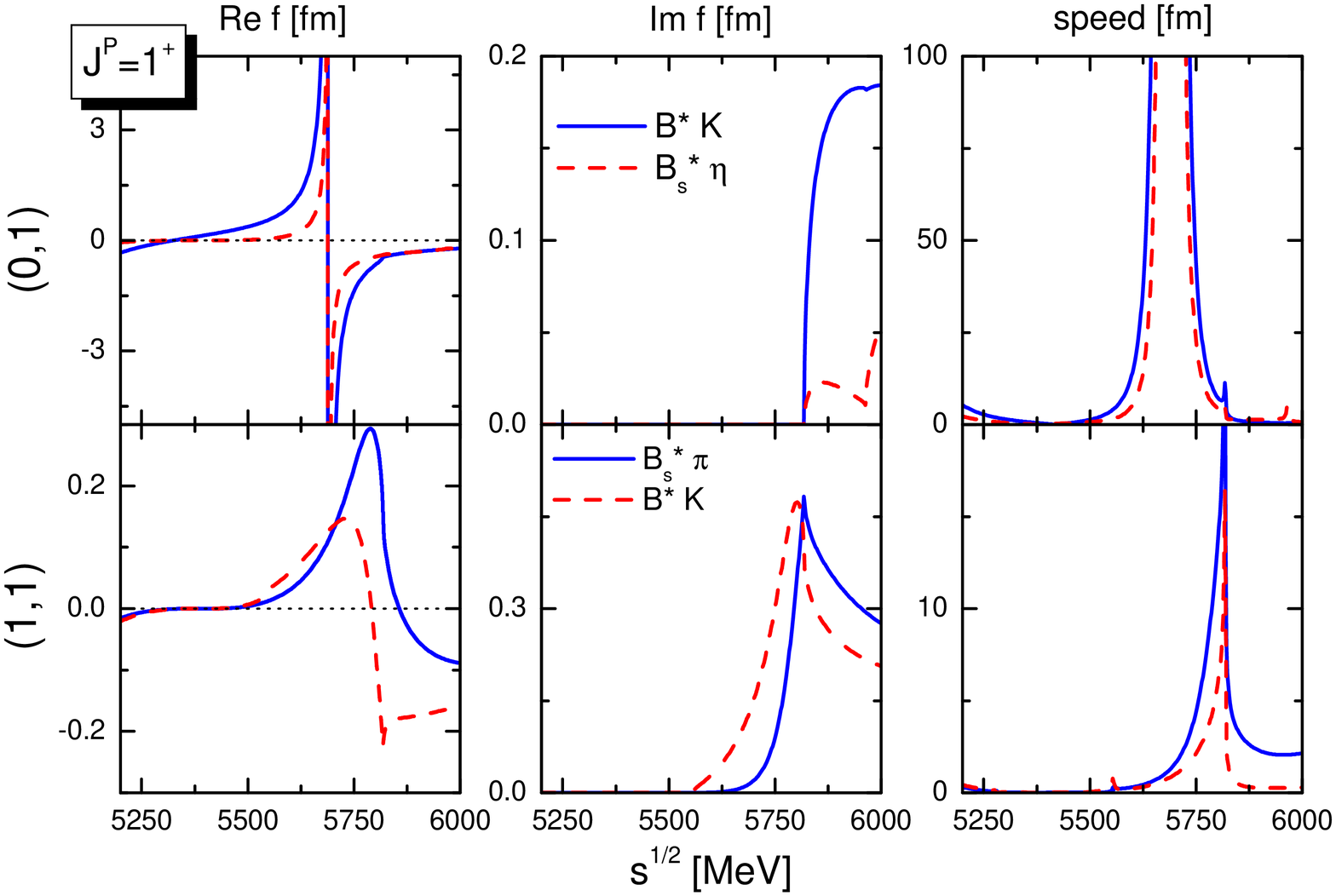}
\end{center}
\caption{Open bottom resonances with $J^P\!=\!0^+,1^+$  and $(I,S)=(0,1),(1,1)$ as seen in
the scattering of Goldstone bosons of $B(5279), B_s(5370)$ and
$B(5325), B_s(5417)$ mesons. Shown speed plots together with real and imaginary parts of reduced scattering amplitude,
$f_{ab}$, with $t_{ab}= f_{ab}\,(p^{(a)}_{\rm cm}\,p^{(b)}_{\rm cm})^{1/2}$ (see (\ref{def-speed})).}\label{fig:3}
\end{figure}

\begin{figure}[t]
\begin{center}
\includegraphics[clip=true,width=13cm]{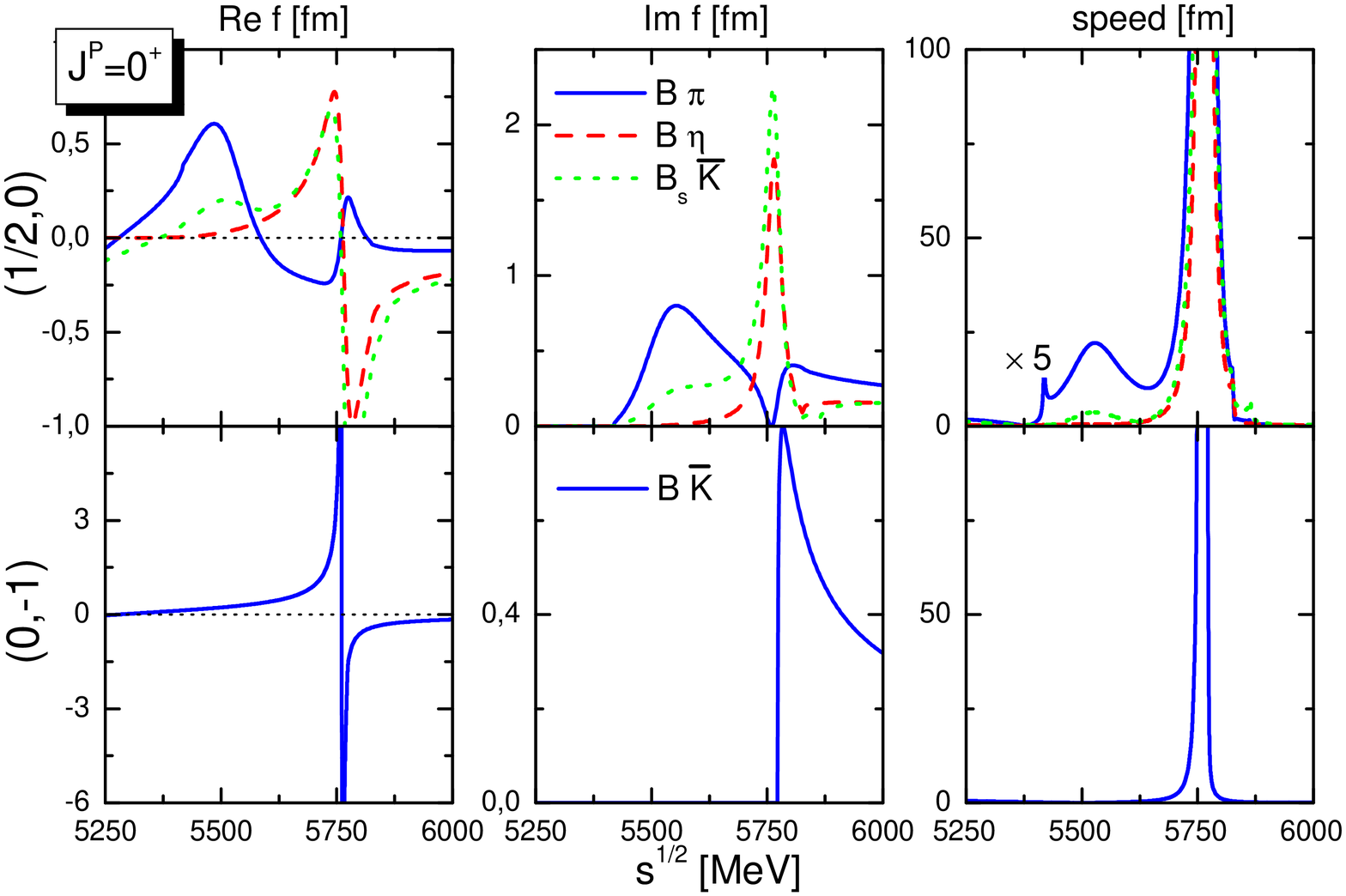}\\
\includegraphics[clip=true,width=13cm]{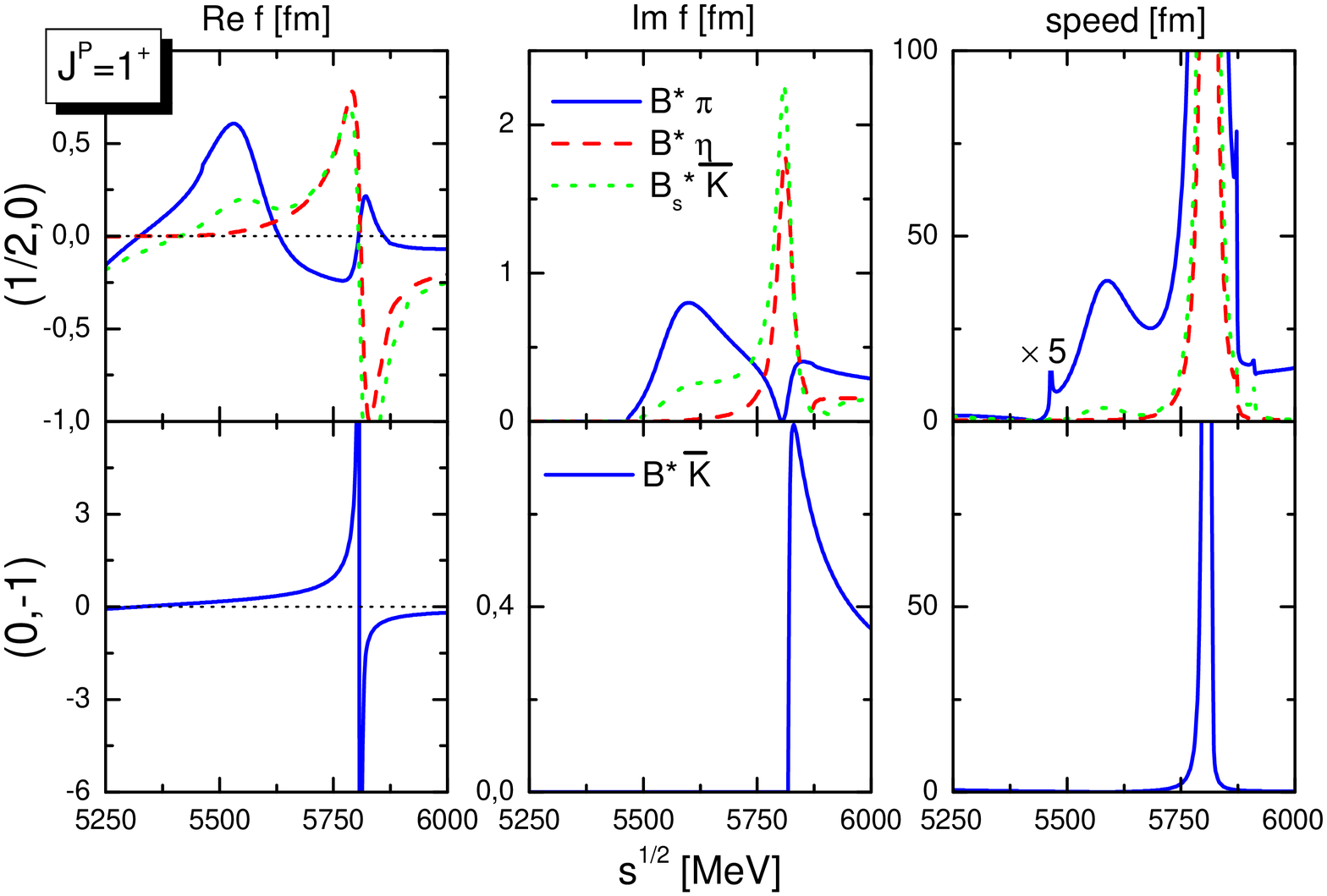}
\end{center}
\caption{Open bottom resonances with $J^P\!=\!0^+,1^+$  and $(I,S)=(\frac{1}{2},0),(0,-1)$ as seen in
the scattering of Goldstone bosons of $B(5279), B_s(5370)$ and
$B(5325), B_s(5417)$ mesons. Shown speed plots together with real and imaginary parts of
reduced scattering amplitude,
$f_{ab}$, with $t_{ab}= f_{ab}\,(p^{(a)}_{\rm cm}\,p^{(b)}_{\rm cm})^{1/2}$
(see (\ref{def-speed})).}\label{fig:4}
\end{figure}

The spectrum predicted for the $1^+$ states is very
similar to the spectrum of the $0^+$ states. Figs. \ref{fig:1},\ref{fig:2} demonstrate that
it is shifted up by approximatively 140 MeV with respect to the $0^+$ spectrum.
The bound state in the $(0,1)$-sector comes at 2440 MeV. Thus the mass splitting of the
$1^+$ and $0^+$ states in this channel agrees very well with the empirical value of about
140 MeV measured by the BABAR and CLEO collaborations \cite{BaBar,CLEO}.
A narrow structure at 2552 MeV is predicted in the $(\frac{1}{2},0)$-channel
which should be identified with the $D(2420)$-resonance \cite{PDG02}. Even though the resonance mass
is overestimated by about 130 MeV our result is consistent with its small width of about
20 MeV. The triplet state in this sector of mass 2325 MeV has again a quite
large width reflecting the strong coupling to the $\pi\,D(2008) $-channel.
Finally we obtain strong cusp effects at the $\bar K \,D(2008)$- and  $K \,D(2008)$-thresholds
in the $(0,-1)$- and $(1,+1)$-sectors. It is interesting to speculate whether chiral correction
terms conspire to slightly increase the net attraction in these sectors. This would lead to a
$(0,-1)$-bound state. The fact that we overestimate the mass of the sextet state
$D(2420)$ by about 130 MeV we take as a strong prediction that this should indeed be the case.
An analogous statement holds for the $0^+$ sector since due to heavy-quark symmetry
chiral correction effects in the $0^+$ and $1^+$ are identical at leading order.

Our predictions for the heavy-light resonance spectrum differ
significantly from the analyses \cite{BEH03,NRZ03} that were based
on the linear realization of the chiral SU(3) symmetry. Besides
the additional sextet states predicted by the more general
non-linear realization of the chiral SU(3) group, most notable are
the differences in the open bottom sector. Using 5279 MeV and 5370
MeV for the $0^-$ ground states \cite{PDG02} we obtain the
following spectrum (see Figs. \ref{fig:3},\ref{fig:4}).
The $(0,1)$ bound-state comes at 5643 MeV, a
value about 70 MeV lower than predicted in \cite{BEH03,NRZ03}. In the
$(1,1)$-channel we predict a broad resonance with mass of about 5750 MeV.
In the $(\frac{1}{2},0)$-channel a
broad state of mass 5526 MeV and a narrow resonance of mass 5760 MeV and
width of about 30 MeV is predicted. Most spectacular is a bound-state just
below the $\bar K\, B$-threshold at 5761 MeV predicted in the $(0,-1)$ channel.

The $1^+$ spectrum we compute in terms of the ground state masses 5325 MeV
and 5417 MeV \cite{PDG02,BEH03}. The results resemble the spectrum
found in the $0^+$ sector. We predict the masses 5690 MeV $(0,1)$, 5790 MeV $(1,1)$,
5590 MeV $(\frac{1}{2},0)$, 5810 MeV
$(\frac{1}{2},0)$, and 5807 MeV $(0,-1)$.

{\bfseries{Acknowledgments}}

M.F.M.L. acknowledges stimulating discussions with M.A. Nowak.

\end{document}